\documentclass{article}
\usepackage{frascatiphys}
\usepackage{graphicx}
\begin{document}
\title{ 
THE FIRST STARS IN THE UNIVERSE
}
\author{
Umberto Maio\\
{\em
INAF -- Osservatorio Astronomico di Trieste, 34143 Trieste (Italy)
}\\
{\em
Leibniz Institute for Astrophysics, 14482 Potsdam (Germany)
}
}
\maketitle
\baselineskip=11.6pt
\begin{abstract}
The basic processes of the formation of the first stars in the primordial Universe are outlined and the implications for cosmological structure formation discussed. By employing theoretical and numerical models of cosmic structure evolution embedded within N-body hydrodynamical chemistry simulations, predictions for the production of the first heavy elements in the Universe are given. These results are then compared against measured data of UV luminosities and metal abundances in different kinds of observations in order to draw conclusions on the chemical and thermal state of the cosmic medium at different cosmological epochs.
\end{abstract}
\baselineskip=14pt

\section{Introduction}
Cosmic structures originate from the growth of matter perturbations at early times in an expanding Universe.
Baryonic objects form from the in-fall and cooling of gas into the dark-matter potential wells since very high redshift ($z$).
A star forming gas `cloud' can form if radiative losses are sufficient to make the gas condense and fragment.
At high $z$, gas cooling is dominated by H, He and H-based molecules, like $\rm H_2$ and HD. After pollution from freshly formed stars, metals also contribute to gas cooling.

Primordial epochs are important for our understanding of the formation of the very first stars and (proto-)galaxies.
These are the sites which witness the occurrence of the first heavy elements in the Universe and their spreading in the surrounding regions.
Unfortunately, there is still a lack of definitive knowledge about the features of the first stars, such as their masses, metal yields and luminosities, so that it is difficult to give exact predictions for cosmic pollution in different environments.
Furthermore, the role of molecules and metals is crucial in dictating the transition from the formation of stars in pristine environments (population III, Pop~III, regime) to a regime which is dominated by cooling due to heavy elements (population II-I, Pop~II-I, regime).

To assess these problems in a quantitative way, deep studies of the hydrodynamical and chemical properties of cosmic medium are required.
Here we will present and summarise results from numerical simulations taking into account the most relevant physical processes to draw conclusions on primordial star formation and on the impacts of the first stars on the following generations of cosmological structures.

\section{Numerical simulations}
For a consistent picture of primordial structure formation it is important to consider a number of processes to be included into N-body hydrodynamical numerical calculations.
We use the cosmological parallel code Gadget\cite{Springel2005} and extend it to include non-equilibrium molecular chemistry, cooling, stellar evolution and metal pollution\cite{Tornatore2004, Maio2007, Maio2010,Petkova2012}.

Molecule formation and evolution determine the first collapsing events through H$_2$ and HD cooling and lead Pop~III star formation.

The occurrence of heavy elements from Pop~III stars increase the efficiency of gas cooling capabilities, hence metal chemistry become important for the formation of second generation stars.

Finally, stellar evolution processes dictate the amounts of photons and heavy elements ejected by stars with different masses and lifetimes via supernova (SN) explosions or asymptotic giant branch (AGB) winds.

The transition\cite{Maio2010,Maio2011} from the Pop~III to the Pop~II-I regime is accounted for according to the underlying metallicity ($Z$) of the collapsing gas.
If star forming gas is pristine or its metallicity is below a threshold limit of $10^{-4}$ the solar value ($Z_\odot$) the actual stellar population is assumed to be Pop~III with a stellar mass range [100, ~500]~M$_\odot$ and an initial mass function (IMF) having a slope of $-2.35$.
If local metallicities are $Z>10^{-4}\,Z_\odot$ then a Pop~II-I regime is assumed with a Salpeter IMF over [0.1, 100]~M$_\odot$.
We stress that the different mass ranges between Pop~III and Pop~II-I generations imply also different metal yields and lifetimes with obvious consequences on the patterns and timescales of cosmological metal enrichment.

From simulated outputs it is possible to extract the main properties of dark-matter haloes, cosmic gas and evolving galaxies and to explore their correlation properties\cite{deSouza2013, Maio2013, deSouza2014, deSouza2015}.

The background cosmology adopted is a model with cold dark matter and cosmological constant $\Lambda$ ($\Lambda$CDM), with 
expansion parameter $\rm H_0 = 70\, km/s/Mpc$,
present total matter density parameter $\Omega_{\rm 0,m} = 0.3$,
baryon density parameter $\Omega_{\rm 0,b} = 0.04$,
$\Lambda$ density parameter $\Omega_{\rm 0,\Lambda} = 0.7$,
mass variance within 8-$\rm Mpc/{\it h}$ radius sphere $\sigma_8 = 0.9$
and spectral index $n=1$.

\section{Results}

\subsection{The first Gyr}
The contribution of the first Pop~III stars to the cosmic star formation rate (SFR) density is shown in Fig.~5 by Maio et al. 2010\cite{Maio2009,Maio2010}.
Given the uncertainties on the critical metallicity for the transition, values of 
$10^{-6}$, $10^{-5}$, $10^{-4}$, $10^{-3}\,\, Z_\odot$ have been tested, as well
(see also Fig.~6 and 7 therein for further parameter dependences).
The contribution from Pop~III regime drops dramatically from about unity at $z>16$ down to $\sim 10^{-3}$ at $z<11$ quite independently from the exact critical metallicity adopted.
This means that at such early epochs ($< 0.5 \,\rm Gyr$) the role of pristine star formation is already marginal and the Universe is already enriched in a sensible way.
The total cosmic star formation results dominated by polluted Pop~II-I haloes, despite their number fraction is not extremely large (below 10 per cent), as shown by Biffi \& Maio (2013)\cite{Biffi2013}, in the top and bottom panel of their Fig.~3.

The picture remains qualitatively similar even after a broader parameter space exploration\cite{Maio2010}. It emerges a primordial Universe which is rapidly enriched and dominated by Pop~II-I stellar population after a relatively short time from the onset of cosmic star formation.

\subsection{Theory and data}
Theoretical results can be compared against available data for luminosity functions (LFs) and specific star formation rates (sSFR).

The LF provides the fraction of objects with a given luminosity (or magnitude) observed at any given redshift.
In the recent years many data for the high-$z$ Universe have become available and they are precious to probe the early epochs at $z>5$.
Results on LFs at such primordial epochs date back to Salvaterra et al.(2013)\cite{Salvaterra2013} where it was found a reasonable agreement between theory and observations for $z>6$ (Fig.~1).
At $z\sim 6$ or lower, the observed LF usually results obscured by effects due to the growth of dust grains in the interstellar medium which need to be taken into account\cite{Dayal2013, MT2015, Mancini2015}.

Data for the sSFR are more uncertain and their scatter is up to one dex, therefore it is quite difficult to disentangle different theoretical models.
Nevertheless, the typical trend of increasing sSFR for increasing $z$ displayed by data is broadly in line with expectations (Fig.~12\cite{Biffi2013}).

\subsection{Implications for high-$z$ GRBs}
Primordial gamma-ray bursts (GRBs) are thought to be originated by the collapse of the first massive stars.
These events can form a black hole and can be accompanied by jets.
Since massive stars are short-lived, early GRBs basically trace star formation episodes and provide information about the local environment.

A number of theoretical studies in this field\cite{Campisi2011, Maio2012, Salvaterra2013, MaioBarkov2014, Ma2015} have shown the potential of GRBs to infer the properties of their host galaxies and place constraints on the primordial Pop~III regime.
It turns out that (Salvaterra et al.; 2013) typical stellar masses of primordial GRB hosts peak at $\sim 10^7\,\rm M_\odot$ and corresponding SFR values range between $\sim 0.01$ and $0.1 \,\rm M_\odot/yr$, giving $\rm sSFR\sim 5-10\, Gyr^{-1}$.
The most popular magnitudes for UV luminosities are in the range [-20, -12], while expected metallicities peak around $Z\sim 10^{-1.5}\,\rm Z_\odot$.
These findings are in agreement with available data at $z>5$, as shown in Fig.~5\cite{Salvaterra2013}, and support the idea that small primordial proto-galaxies produce most of the ionising photons at early times (Fig.~3-4 therein).

Further analyses\cite{Ma2015} have stressed the most suited targets for Pop~III searches at higher $z$.
In particular, Ma et al. (2015) have explored the indirect Pop~III signatures imprinted in enriched Pop~II-I hosts in order to disentangle the two stellar populations. Thus, Pop~II-I star forming galaxies pre-enriched by very massive Pop~III stars appear to have typical metallicities $Z<10^{-2.8}\,\rm Z_\odot$ and peculiar abundance ratios, such as $\rm [Si/O]<-0.6$, $\rm [S/O]<-0.6$, $\rm [C/O]>-0.4$.
Obviously, such criteria depend on the assumed mass of primordial stars, hence, spectral data can give additional hints on the stellar structure of such objects.

\section{Conclusions and perspectives}
The results presented here have been obtained by {\it ad hoc} numerical simulations including N-body and hydrodynamical calculations, atomic and molecular chemistry, star formation, stellar evolution and feedback effects.

First star formation episodes are very `bursty' and efficient metal spreading leads to a rapid transition from Pop~III to Pop~II-I regime\cite{Maio2010}.

Primordial Pop~III stars dominate cosmic star formation for a relatively short time and only a residual fraction of SFR survives at late times.
This makes direct searches of Pop~III stars statistically very difficult and implies the need to rely on additional indirect methods to shade light on primordial star forming events.

Among the possible processes that could affect the final results it is worth mentioning the possible existence of primordial supersonic flows originated at decoupling.
They could induce homogeneous gas streaming motions on Mpc scales and have impacts on early structure formation, reionization and the lowest-mass dwarf galaxies\cite{Maio2011bf, Maio2015ska}.

Results, though, are not very sensitive to the assumed values for the critical metallicity, Pop~III metal yields or IMF slope.

In these kinds of studies the commonly assumed background framework is the $\Lambda$CDM model, however different scenarios are possible.
Alternative non-Gaussian models\cite{MaioIannuzzi2011, Maio2011cqg, MK2012, Maio2012, Pace2014} and dark-energy quintessence models\cite{Maio2006} have been tested, as well.
Despite the details in these cases might change the overall trends are generally recovered.

Also the elementary nature of dark matter might play a role leaving room for warm dark matter in place of cold dark matter\cite{MV2015}.
The dumping effects of warm dark matter at small scales seem to be evident in terms of dark-matter structure distributions and shapes, however the implications on the first stars and reionization are less trivial.

Some final considerations are briefly devoted to future perspectives to observe the first stars and galaxies in the next decades.
The upcoming James Webb Space Telescope (JWST)\footnote{
http://www.jwst.nasa.gov
}
is under construction and has been designed to detect faint sources at early times and to study reionization and early galaxy formation.
The Square Kilometer Array (SKA)\footnote{
https://www.skatelescope.org
} 
radio telescope will be built in the South hemisphere and will give scientists the possibility to study a large number of topics, among which hydrogen distribution, galaxy formation and radio emission in the first billion years\cite{Koopmans2015}.
The Athena\footnote{
http://www.the-athena-x-ray-observatory.eu
}
space mission will target the hot and energetic Universe by employing high-resolution X-ray spectroscopy. Its capabilities will be exploited to investigate also GRB X-ray afterglows up to redshift $z\sim 6-10$.

In spite of the huge costs for the technical development of such experiments, the scientific return is supposed to be unprecedented.

\newcommand{\mnras}{Mon. Not. R. Astron. Soc.}
\newcommand{\aap}{A\&A}

\end{document}